\documentclass[twocolumn,showpacs,preprintnumbers,amsmath,amssymb]{revtex4-1}
\usepackage{comment}
\usepackage{graphicx,graphics,wrapfig,rotating}         
\usepackage{epsfig,epstopdf, subfigure}
\usepackage{dcolumn}                                    
\usepackage{bm,fancybox}
\usepackage{amsmath,amssymb}                          
\usepackage{times,euscript,oldgerm}              %
\usepackage[english]{babel}                      %
\usepackage{psfrag}
\usepackage{hyperref}
\usepackage{multibib}
\usepackage{bbold}
\newcommand{\beq}{\begin{equation}}
\newcommand{\eeq}{\end{equation}}
\newcommand{\beqnarray}{\begin{eqnarray}}
\newcommand{\eeqnarray}{\end{eqnarray}}
\newcommand{\Renyi}{R\'{e}nyi\hspace{1mm}}
\newcommand{\tr}[1]{{\rm{Tr}}{#1}}
\begin{document}
\title{Measuring entanglement entropy through the interference of quantum many-body twins}
\author{Rajibul Islam, Ruichao Ma, Philipp M. Preiss, M. Eric Tai, Alexander Lukin, Matthew Rispoli, Markus Greiner}
\affiliation{Department of Physics, Harvard University, Cambridge, MA 02138, USA}

 \begin{abstract}
Entanglement is one of the most intriguing features of quantum mechanics. It describes non-local correlations between quantum objects, and is at the heart of quantum information sciences. 
Entanglement is rapidly gaining prominence in diverse fields ranging from condensed matter to quantum gravity. Despite this generality, measuring entanglement remains challenging. This is especially true in systems of interacting delocalized particles, for which a direct experimental measurement of spatial entanglement has been elusive. Here, we measure entanglement in such a system of itinerant particles using quantum interference of many-body twins. Leveraging our single-site resolved control of ultra-cold bosonic atoms in optical lattices, we prepare and interfere two identical copies of a many-body state. This enables us to directly measure quantum purity,  \Renyi entanglement entropy, and mutual information. These experiments pave the way for using entanglement to characterize quantum phases and dynamics of strongly-correlated many-body systems.
 \end{abstract}
\date{\today}
\maketitle

At the heart of quantum mechanics lies the principle of superposition: a quantum system can be in several states at the same time. Measurement on such a superposition state will exhibit randomness in the outcomes. This quantum randomness is fundamental in nature, unlike classical randomness that arises when the observer has incomplete knowledge or ignores information about the system, as when throwing dice or flipping coins. In a many-body quantum system, quantum superposition between various possible configurations often results in a correlated randomness in the measurement outcomes of different parts of the system. These correlated subsystems are then said to be entangled \cite{Horodecki2009}. The non-local correlations between entangled subsystems prompted Einstein to describe entanglement as `spooky action at a distance' \cite{Einstein1935}, and were shown by Bell to be inconsistent with reasonable local theories of classical hidden variables \cite{Bell1964}. Later, it was realized that entanglement could be used as a resource to perform tasks not possible classically, with applications in computation \cite{Shor1997, Nielsen2010}, communication \cite{Bennett1993}, and simulating the physics of strongly correlated quantum systems \cite{Feynman1982}.

In few level quantum systems, entangled states have been investigated extensively for studying the foundations of quantum mechanics \cite{Aspect1999} and as a resource for quantum information applications \cite{Nielsen2010, Ladd2010}. Recently, it was realized that the concept of entanglement has broad impact in many areas of quantum many-body physics, ranging from condensed matter \cite{Amico2008} to high energy field theory \cite{Calabrese2009} and quantum gravity \cite{Nishioka2009}. In this general context, entanglement is most often quantified by the entropy of entanglement \cite{Horodecki2009} that arises in a subsystem when the information about the remaining system is ignored. This entanglement entropy exhibits qualitatively different behavior than classical entropy and has been used in theoretical physics to probe various properties of the many-body system. In condensed matter physics, for example, the scaling behavior \cite{Eisert2010} of entanglement entropy allows distinguishing between phases that cannot be characterized by symmetry properties, such as topological states of matter \cite{Kitaev2006, Levin2006, Jiang2012} and spin liquids \cite{Zhang2011, Isakov2011}. Entanglement entropy can be used to probe quantum criticality \cite{Vidal2003} and non-equilibrium dynamics \cite{Bardarson2012, Daley2012}, and to determine whether efficient numerical techniques for computing many-body physics exist \cite{Schuch2008}. 
\begin{figure}
\begin{center}
\includegraphics[width=0.9 \linewidth]{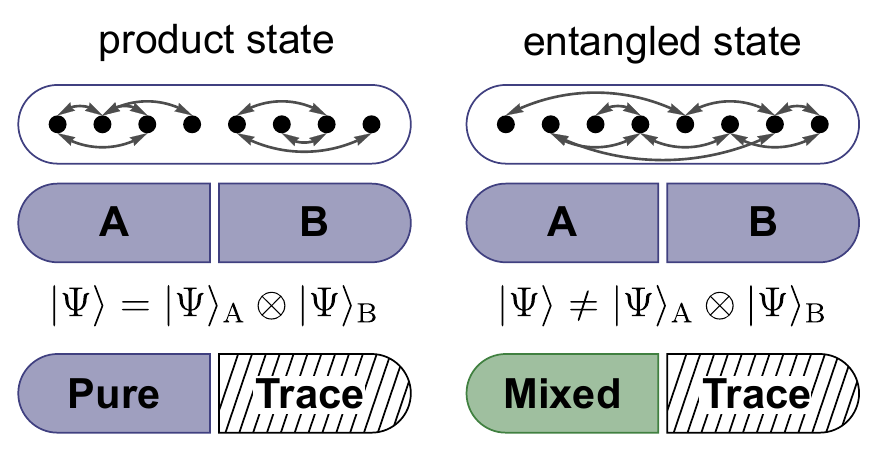}
\end{center}
\caption{\label{fig:partial}
{{\bf Bipartite entanglement and partial measurements.} A generic pure quantum many-body state has quantum correlations (shown as arrows) between different parts. If the system is divided into two subsystems A and B, the subsystems will be bipartite entangled with each other when quantum correlations span between them (right column). Only with no bipartite entanglement present, the partitioned system $|\psi_{AB}\rangle$ can be described as a product of subsystem states $|\psi_{A}\rangle$ and $|\psi_{B}\rangle$ (left column). A path for measuring the bipartite entanglement emerges from the concept of partial measurements: ignoring all information about subsystem B (indicated as ``Trace'') will put subsystem A into a statistical mixture, to a degree given by the amount of bipartite entanglement present. Finding ways of measuring the many-body quantum state purity of the system and comparing that of its subsystems would then enable measurements of entanglement. For an entangled state, the subsystems will have less purity than the full system.}}
\end{figure}
Despite the growing importance of entanglement in theoretical physics, current condensed matter experiments do not have a direct probe to detect and measure entanglement. Synthetic quantum systems such as cold atoms \cite{Bloch2012, Blatt2012}, photonic networks \cite{Aspuru2012}, and some microscopic solid state devices \cite{Houck2012} have unique advantages: their almost arbitrary control and detection of single particles, experimental access to relevant dynamical time scales, and isolation from the environment. In these systems, specific entangled states of few qubits, such as the highly entangled Greenberger-Horne-Zeilinger (GHZ) state \cite{Bouwmeester1999} have been experimentally created and detected using witness operators \cite{Guhne2009}. However, entanglement witnesses are state specific. An exhaustive method to measure entanglement of an arbitrary state requires reconstructing the quantum state using tomography \cite{James2001}. This has been accomplished in small systems of photonic qubits \cite{Pan2012} and trapped ion spins \cite{Haffner2005}, but there is no known scheme to perform tomography for systems involving itinerant delocalized particles. With multiple copies of a system, however, one can use quantum many-body interference to quantify entanglement even in itinerant systems \cite{Ekert2002, Alves2004, Daley2012}.

In this work, we take advantage of the precise control and readout  afforded by our quantum gas microscope \cite{Bakr2010} to prepare and interfere two identical copies of a four-site Bose-Hubbard system. This many-body quantum interference enables us to measure quantities that are not directly accessible in a single system, e.g.  quadratic functions of the density matrix \cite{Ekert2002, Alves2004, Brun2004, Daley2012, Bovino2005, Walborn2006, Schmid2008}. Such non-linear functions can reveal entanglement \cite{Horodecki2009}. In our system, we directly measure the quantum purity, \Renyi entanglement entropy, and mutual information to probe the entanglement in site occupation numbers.

\section*{Bipartite entanglement}
To detect entanglement in our system, we use a fundamental property of entanglement between two subsystems (bipartite entanglement): ignoring information about one subsystem results in the other becoming a classical mixture of pure quantum states. This classical mixture in a density matrix $\rho$ can be quantified by measuring the quantum purity, defined as ${\rm Tr}(\rho^{2})$. For a pure quantum state the density matrix is a projector and ${\rm Tr}(\rho^{2})=1$, whereas for a mixed state ${\rm Tr}(\rho^{2})<1$. In case of a product state, the subsystems $A$ and $B$ of a many-body system $AB$ described by a wavefunction $|\psi_{AB}\rangle$ (Fig. \ref{fig:partial}) are individually pure as well, i.e. ${\rm Tr} (\rho_{A}^{2})={\rm Tr} (\rho_{B}^{2})={\rm Tr} (\rho_{AB}^{2})=1$. Here the reduced density matrix of $A$, $\rho_{A}={\rm Tr}_{B} (\rho_{AB})$, where $\rho_{AB}=|\psi_{AB}\rangle\langle\psi_{AB}|$ is the density matrix of the full system. ${\rm Tr}_{B}$ indicates tracing over or ignoring all information about the subsystem $B$. For an entangled state, the subsystems become less pure compared to the full system as the correlations between $A$ and $B$ are ignored in the reduced density matrix, ${\rm Tr} (\rho_{A}^{2})={\rm Tr} (\rho_{B}^{2})<{\rm Tr} (\rho_{AB}^{2})=1$. Even if the many-body state is mixed (${\rm Tr}(\rho_{AB}^{2})<1$), it is still possible to measure entanglement between the subsystems \cite{Horodecki2009}. It is sufficient \cite{Horodecki1996} to prove this entanglement by showing that the subsystems are less pure than the full system, i.e.
\begin{eqnarray}
{\rm Tr} (\rho_{A}^{2})<{\rm Tr} (\rho_{AB}^{2}),\nonumber\\
{\rm Tr} (\rho_{B}^{2})<{\rm Tr} (\rho_{AB}^{2}).
\label{Inequality}
\end{eqnarray}
These inequalities provide a powerful tool for detecting entanglement in the presence of experimental imperfections. Furthermore, quantitative bounds on the entanglement present in a mixed many-body state can be obtained from these state purities \cite{Mintert2007}. 

Eq.(\ref{Inequality}) can be framed in terms of entropic quantities \cite{Horodecki2009, Horodecki1996}. A particularly useful and well studied quantity is the $n$-th order \Renyi entropy, 
\beq
S_{n}(A)=\frac{1}{1-n}\log{\tr{(\rho_{A}^n)}}.
\label{Renyi}
\eeq 
From Eq. (\ref{Renyi}), we see that the second-order ($n=2$) \Renyi entropy and purity are related by ${S_{2}(A)=-\log{{\rm{Tr}}(\rho_{A}^{2})}}$. $S_{2}(A)$ provides a lower bound for the von Neumann entanglement entropy ${S_{VN}(A)=S_{1}(A)=-\tr{(\rho_{A}\log{\rho_{A}})}}$ extensively studied theoretically. The \Renyi entropies are rapidly gaining importance in theoretical condensed matter physics, as they can be used to extract information about the ``entanglement spectrum'' \cite{Hui2008} providing more complete information about the quantum state than just the von Neuman entropy. In terms of the second-order \Renyi entropy, the sufficient conditions to demonstrate entanglement \cite{Horodecki1996, Horodecki2009} become $S_{2}(A)>S_{2}(AB)$, and $S_{2}(B)>S_{2}(AB)$, i.e. the subsystems have more entropy than the full system. These entropic inequalities are more powerful in detecting certain entangled states than other inequalities like the Clauser-Horne-Shimony-Holt (CHSH) inequality \cite{Bovino2005, Horodecki1996}.

\begin{figure}
\begin{center}
\includegraphics[width=0.95 \linewidth]{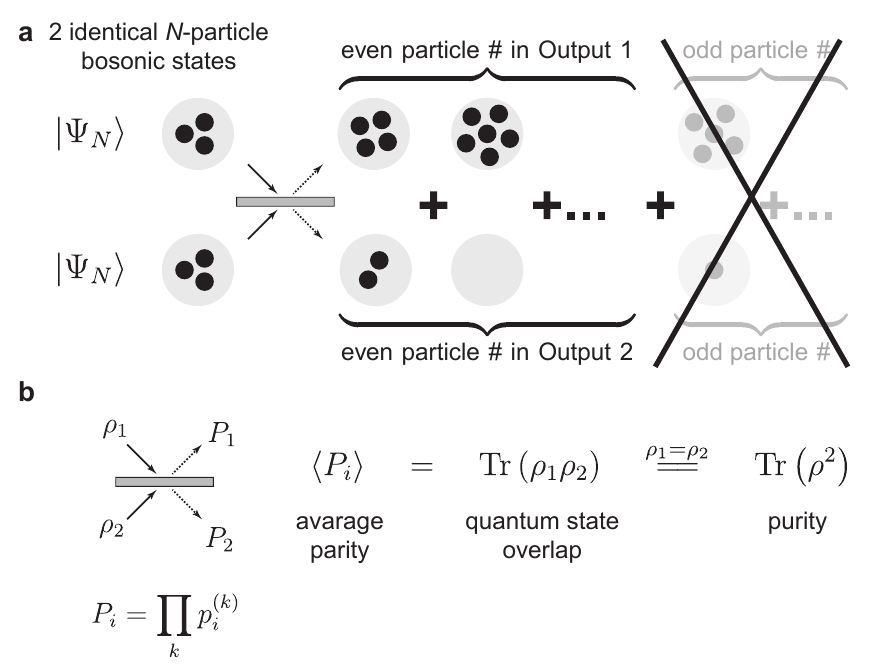}
\end{center}
\caption{\label{fig:manybodyBS}
{{\bf Measurement of quantum purity with many-body bosonic interference of quantum twins.} {\bf a.} When two $N$ particle bosonic systems that are in identical pure quantum states are interfered on a 50\%-50\% beam splitter, they always produce output states with even number of particles in each copy. This is due to the destructive interference of odd outcomes and represents a generalized Hong-Ou-Mandel interference, in which two identical photons always appear in pairs after interfering on a beam splitter. {\bf b.} If the input states $\rho_{1}$ and $\rho_{2}$ are not perfectly identical or not perfectly pure, the interference contrast is reduced. In this case the expectation value of the parity of particle number $\langle P_{i}\rangle$ in output $i=1,2$ measures the quantum state overlap between the two input states. For two identical input states $\rho_{1}=\rho_{2}$, the average parity $\langle P_{i}\rangle$ therefore directly measures the quantum purity of the states. We only assume that the input states have no relative macroscopic phase relationship.}}
\end{figure}

\section*{Measurement of quantum purity}
The quantum purity and hence the second-order \Renyi entropy can be directly measured by interfering two identical and independent copies of the quantum state on a 50\%-50\% beam splitter \cite{Daley2012, Ekert2002, Alves2004, Bovino2005}. For two identical copies of a bosonic Fock state, the output ports always have even particle number, as illustrated in Fig. \ref{fig:manybodyBS}a. This is due to the destructive interference of all odd outcomes. If the system is composed of multiple modes, such as internal spin states or various lattice sites, the total number parity $P_{i}=\prod_{k}p_{i}^{(k)}$ is equal to unity in the output ports $i=1,2$. Here the parity for mode $k$, $p_{i}^{(k)}=\pm 1$ for even or odd number of particles, respectively. The well known Hong-Ou-Mandel (HOM) interference of two identical single photons \cite{Hong1987} is a special case of this scenario. Here a pair of indistinguishable photons incident upon different input ports of a 50\%-50\% beam splitter undergoes bosonic interference such that both photons always exit from the same output port. In general, the average parity measured in the many-body bosonic interference on a beam splitter probes the quantum state overlap between the two copies $\langle P_{i}\rangle={\rm Tr} (\rho_{1}\rho_{2})$, where $\rho_{1}$ and $\rho_{2}$ are the density matrices of the two copies respectively and $\langle ...\rangle$ denotes averaging over repeated experimental realizations or over identical systems, as shown in Fig. \ref{fig:manybodyBS}b. Hence, for two identical systems, i.e. for $\rho_{1}=\rho_{2}=\rho$, the average parity for both output ports ($i=1,2$) equals the quantum purity of the many-body state \cite{Ekert2002, Alves2004, Daley2012}, 
\begin{equation}
\label{ParityPurity}
\langle P_{i}\rangle = {\rm Tr}(\rho^{2}).
\end{equation}
Equation (\ref{ParityPurity}) represents the most important theoretical foundation behind this work -- it connects a quantity depending on quantum coherences in the system to a simple observable in the number of particles. It holds even without fixed particle number, as long as there is no definite phase relationship between the copies (Supplementary material). From Eqs. (\ref{Inequality}) and (\ref{ParityPurity}), detecting entanglement in an experiment reduces to measuring the average particle number parity in the output ports of the multi-mode beam splitter.

\begin{figure}
\begin{center}
\includegraphics[width=1 \linewidth]{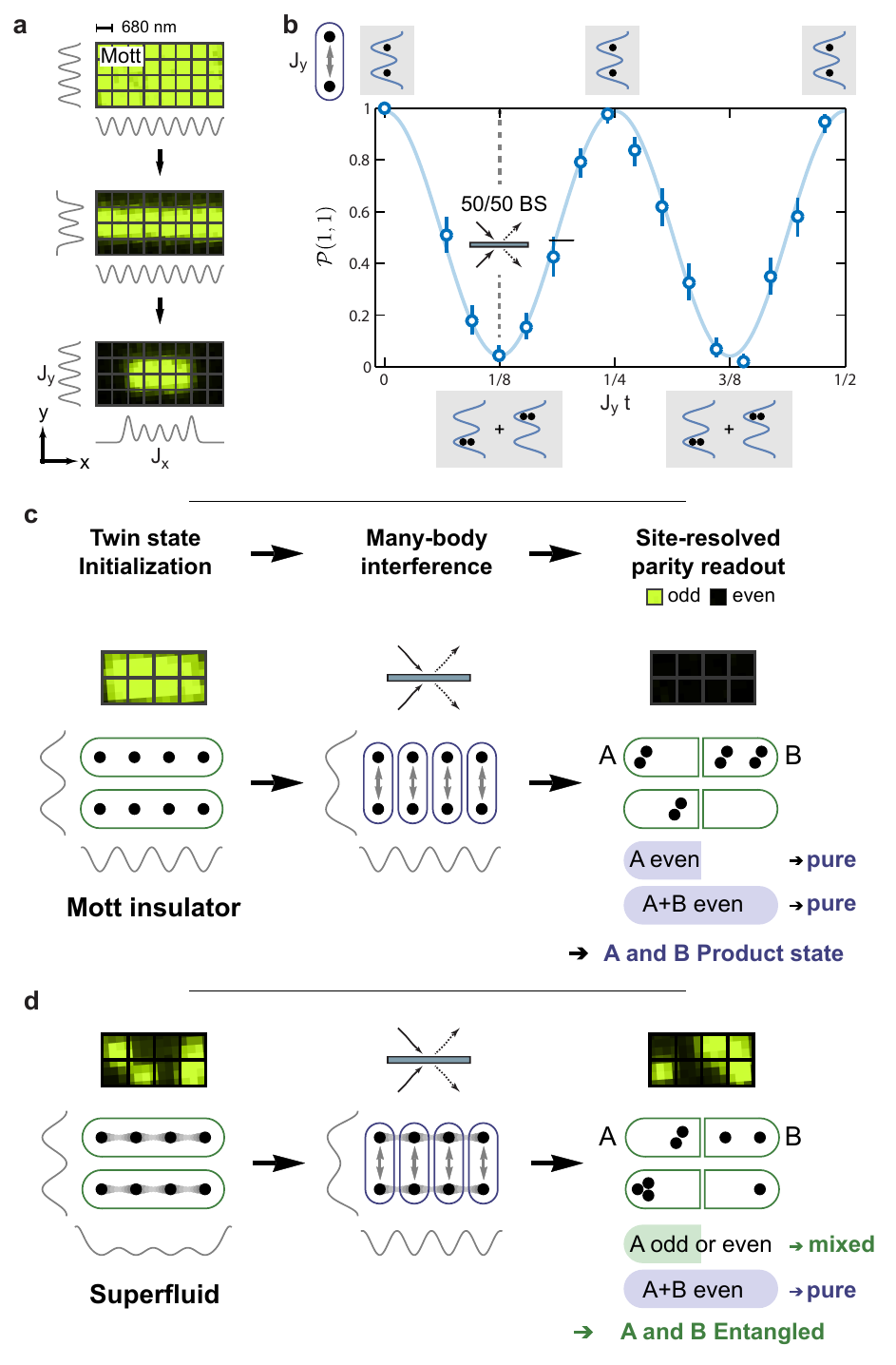}
\end{center}
\caption{\label{fig:HOM}
{{\bf Many-body interference to probe entanglement in optical lattices.} {\bf a.} A high resolution microscope is used to directly image the number parity of ultra cold bosonic atoms on each lattice site (raw images: green = odd, black = even). Two adjacent 1D lattices are created by combining an optical lattice and potentials created by a spatial light modulator (SLM). We initialize two identical many-body states by filling the potentials from a low entropy 2D Mott insulator. The tunneling rates $J_{x}$, $J_{y}$ can be tuned independently by changing the depth of the potential. {\bf b.} The atomic beam splitter operation is realized in a tunnel coupled double well potential. An atom, initially localized in one of the wells, delocalizes with equal probability into both the wells by this beam splitter. Here, we show the atomic analog of the HOM interference of two states. The joint probability $\mathcal{P}(1,1)$ measures the probability of coincidence detection of the atoms in separate wells as a function of normalized tunnel time $J_{y}t$, with the single particle tunneling $J_{y}=193(4)$ Hz. At the beam splitter duration ($J_{y}t=1/8$) bosonic interference leads to a nearly vanishing $\mathcal{P}(1,1)$ corresponding to an even parity in the output states. This can be interpreted as a measurement of the purity of the initial Fock state, here measured to be $\approx 0.90(4)$. The data shown here are averaged over two independent double wells. The blue curve is a maximum likelihood fit to the data, and the errorbars reflect 1 $\sigma$ statistical error. {\bf c.} When two copies of a product state, such as the Mott insulator in the atomic limit are interfered on the beam splitter, the output states contain even number of particles globally (full system) as well as locally (subsystem), indicating pure states in both. {\bf d.} On the other hand, for two copies of an entangled state, such as a superfluid state, the output states contain even number of particles globally (pure state) but a mixture of odd and even outcomes locally (mixed state). This directly demonstrates entanglement. }}
\end{figure}

We probe entanglement formation in a system of interacting $^{87}$Rb atoms on a one dimensional optical lattice with a lattice constant of 680 nm. The dynamics of atoms in the lattice is described by the Bose-Hubbard Hamiltonian,
\begin{equation}
H=-J\sum_{\langle i,j\rangle}a^{\dag}_{i}a_{j} + \frac{U}{2}\sum_{i}n_{i}(n_{i}-1),
\label{eqn:H}
\end{equation}
where $a^{\dag}_{i}$, $a_{i}$ and $n_{i}=a^{\dag}_{i}a_{i}$ are the bosonic creation and annihilation operators, and the number of atoms at site $i$, respectively. The atoms tunnel between neighboring lattice sites (indicated by $\langle i,j\rangle$) with a rate $J$ and experience an onsite repulsive interaction energy $U$. The Planck's constant $h$ is set to 1 and hence both $J$ and $U$ are expressed in Hz. The dimensionless parameter $U/J$ is controlled by the depth of the optical lattice. Additionally, we can superimpose an arbitrary optical potential with a resolution of a single lattice site by using a spatial light modulator (SLM) as an amplitude hologram through a high resolution microscope (Supplementary material). This microscope also allows us to image the number parity of each lattice site independently \cite{Bakr2010}. 

To initialize two independent and identical copies of a state with fixed particle number $N$,  we start with a low entropy 2D Mott insulator with unity filling in the atomic limit \cite{Bakr2010} and deterministically retain a plaquette of $2\times N$ atoms while removing all others (Supplementary material). This is illustrated in Fig. \ref{fig:HOM}a. The plaquette of $2\times N$ atoms contains two copies (along the $y$-direction) of an $N$-atom one-dimensional system (along the $x$-direction), with $N=4$ in this figure. The desired quantum state is prepared by manipulating the depth of the optical lattice along $x$, varying the parameter $U/J_{x}$ where $J_{x}$ is the tunneling rate along $x$. A box potential created by the SLM is superimposed onto this optical lattice to constrain the dynamics to the sites within each copy. During the state preparation, a deep lattice barrier separates the two copies and makes them independent of each other.

The beam splitter operation required for the many-body interference is realized in a double well potential along $y$. The dynamics of atoms in the double well is likewise described by the Bose-Hubbard hamiltonian, Eq. (\ref{eqn:H}). A single atom, initially localized in one well, undergoes coherent Rabi oscillation between the wells with a Rabi frequency of $J=J_{y}$ (oscillation frequency in the amplitude). At discrete times during this evolution, $t=t_{BS}=\frac{1}{8J_{y}}, \frac{3}{8J_{y}},...$, the atom is delocalized equally over the two wells with a fixed phase relationship. Each of these times realizes a beam splitter operation, for which the same two wells serve as the input ports at time $t=0$ and output ports at time $t=t_{BS}$. Two indistinguishable atoms with negligible interaction strength ($U/J_{y}\ll 1$) in this double well will undergo interference as they tunnel. The dynamics of two atoms in the double well is demonstrated in Fig. \ref{fig:HOM}b  in terms of the joint probability  $\mathcal{P}(1,1)$ of finding them in separate wells versus the normalized time $J_{y}t$. The joint probability $\mathcal{P}(1,1)$ oscillates at a frequency of 772(16) Hz $= 4 J_{y}$, with a contrast of about 95(3)\%. At the beam splitter times, $\mathcal{P}(1,1)\approx 0$. The first beam splitter time, $tJ_{y}=\frac{1}{8}$ is used for all the following experiments, with $\mathcal{P}(1,1)\approx 0.05(2)$. This is a signature of bosonic interference of two indistinguishable particles \cite{Kaufman2014, Lopes2015} akin to the photonic HOM interference \cite{Hong1987}. This high interference contrast indicates the near-perfect suppression of classical noise and fluctuations and includes an expected 0.6\% reduction due to finite interaction strength ($U/J_{y}\approx 0.3$). The results from this interference can be interpreted as a measurement of the quantum purity of the initial Fock state as measured from the average parity (Eq.(\ref{ParityPurity})), $\langle P_{i}\rangle=1-2\times \mathcal{P}(1,1)=0.90(4)$, where $i=1,2$ are the two copies. 
\begin{figure*}
\begin{flushleft}
\includegraphics[width=1 \linewidth]{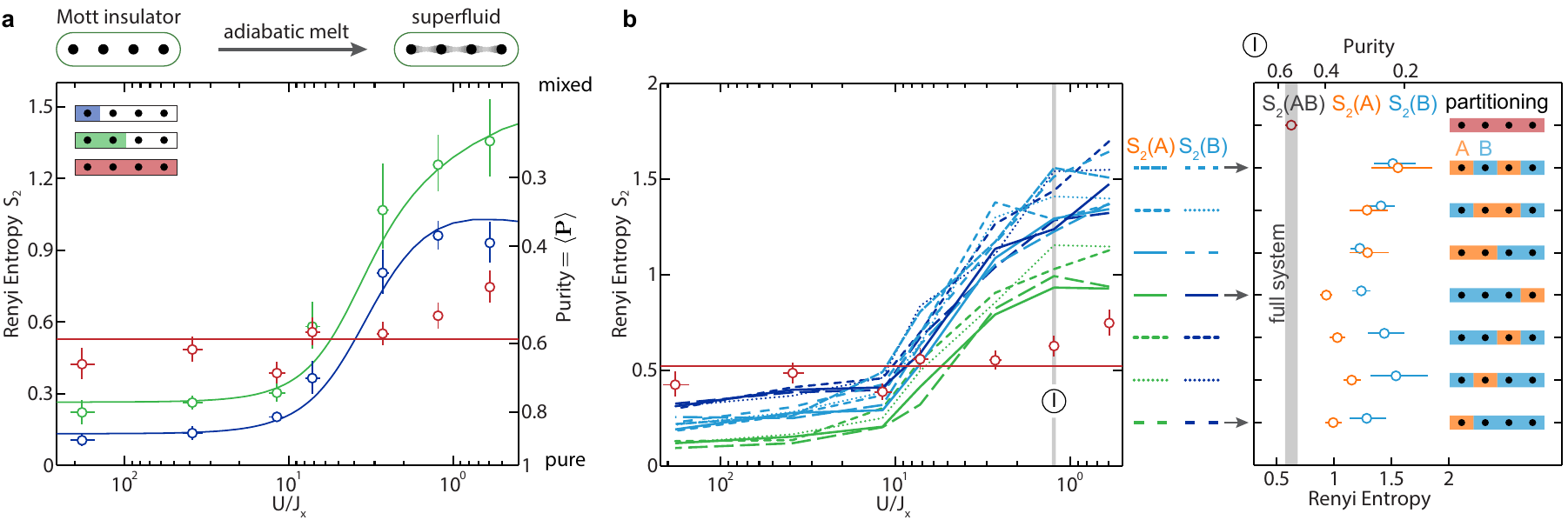}
\end{flushleft}

\caption{\label{fig:S2}
{{\bf Entanglement in the ground state of the Bose-Hubbard model.} We study the Mott insulator to superfluid transition with four atoms on four lattice sites in the ground state of the Bose-Hubbard model, Eq. (\ref{eqn:H}). {\textbf a.} As the interaction strength $U/J_{x}$ is adiabatically reduced the purity of the subsystem $A$ (green and blue, inset), ${\rm Tr}(\rho_{A}^{2})$, become less than that of the full system (red). This demonstrates entanglement in the superfluid phase, generated by coherent tunneling of bosons across lattice sites. In terms of the second-order \Renyi entanglement entropy, $S_{2}(A)=-\log{{\rm Tr}(\rho_{A}^{2})}$, the full system has less entropy than its subsystems in this state. In the Mott insulator phase ($U/J_{x}\gg 1$) the full system has more \Renyi entropy (and less purity) than the subsystems, due to the lack of sufficient entanglement and a contribution of classical entropy. The circles are data and the solid lines are theory calculated from exact diagonalization. The only free parameter is an added offset, assumed extensive in system size and consistent with the average measured entropy in the full system. {\textbf b.} Second-order \Renyi entropy of all possible bi-partitioning of the system. For small $U/J_{x}$, all subsystems (data points connected by green and blue lines) have more entropy than the full system (red circles), indicating full multipartite entanglement \cite{Palmer2005} between the four lattice sites. The residual entropy in the Mott insulating regime is from classical entropy in the experiment, and extensive in the subsystem size. {\textit Right:} The values of all Renyi entropies of the particular case of $U/J_{x}\approx 1$ are plotted, to demonstrate spatial multipartite entanglement in this superfluid.}}
\end{figure*}

\section*{Entanglement in the ground state}

The Bose-Hubbard model provides an interesting system to investigate entanglement. In optical lattice systems, a lower bound of the spatial entanglement has been previously estimated from time-of-flight measurements \cite{Cramer2013} and entanglement dynamics in spin degrees-of-freedom has been investigated with partial state reconstruction \cite{Fukuhara2015}. Here, we directly measure entanglement in real space occupational particle number in a site-resolved way. In the strongly interacting, atomic limit of $U/J_{x}\gg 1$, the ground state is a Mott insulator corresponding to a Fock state of one atom at each lattice site. The quantum state has no spatial entanglement with respect to any partitioning in this phase -- it is in a product state of the Fock states. As the interaction strength is reduced adiabatically, atoms begin to tunnel across the lattice sites, and ultimately the Mott insulator melts into a superfluid with a fixed atom number. The delocalization of atoms create entanglement between spatial subsystems. This entanglement originates \cite{Verstraete2003, Bartlett2003, Schuch2004} from correlated fluctuations in the number of particles between the subsystems due to the super-selection rule that the total particle number in the full system is fixed, as well as coherence between various configurations without any such fluctuation.

To probe the emergence of entanglement, we first prepare the ground state of Eq. (\ref{eqn:H}) in both the copies by adiabatically lowering the optical lattice potential along $x$. Then we freeze the tunneling along $x$ without destroying the coherence in the many-body state and apply the beam splitter along $y$. Finally, we rapidly turn on a very deep 2D lattice to suppress all tunneling and detect the atom number parity (even = 1, odd = -1) at each site. We construct the parity of a spatial region by multiplying the parities of all the sites within that region. The average parity over repeated realizations measures the quantum purity, both globally and locally according to Eq. (\ref{ParityPurity}), enabling us to determine the second-order \Renyi entropy globally and for all possible subsystems. In the atomic Mott limit (Fig. \ref{fig:HOM}c), the state is separable. Hence, the interference signal between two copies should show even parity in all subsystems, indicating a pure state with zero entanglement entropy. Towards the superfluid regime (Fig. \ref{fig:HOM}d), the buildup of entanglement leads to mixed states 
in subsystems, corresponding to a finite entanglement entropy. Hence, the measurement outcomes do not have a pre-determined parity. Remarkably, the outcomes should still retain even global parity, indicating a pure global state. Higher entropy in the subsystems than the global system cannot be explained classically and demonstrates bipartite entanglement.

\begin{figure*}
\begin{flushleft}
\includegraphics[width=1 \linewidth]{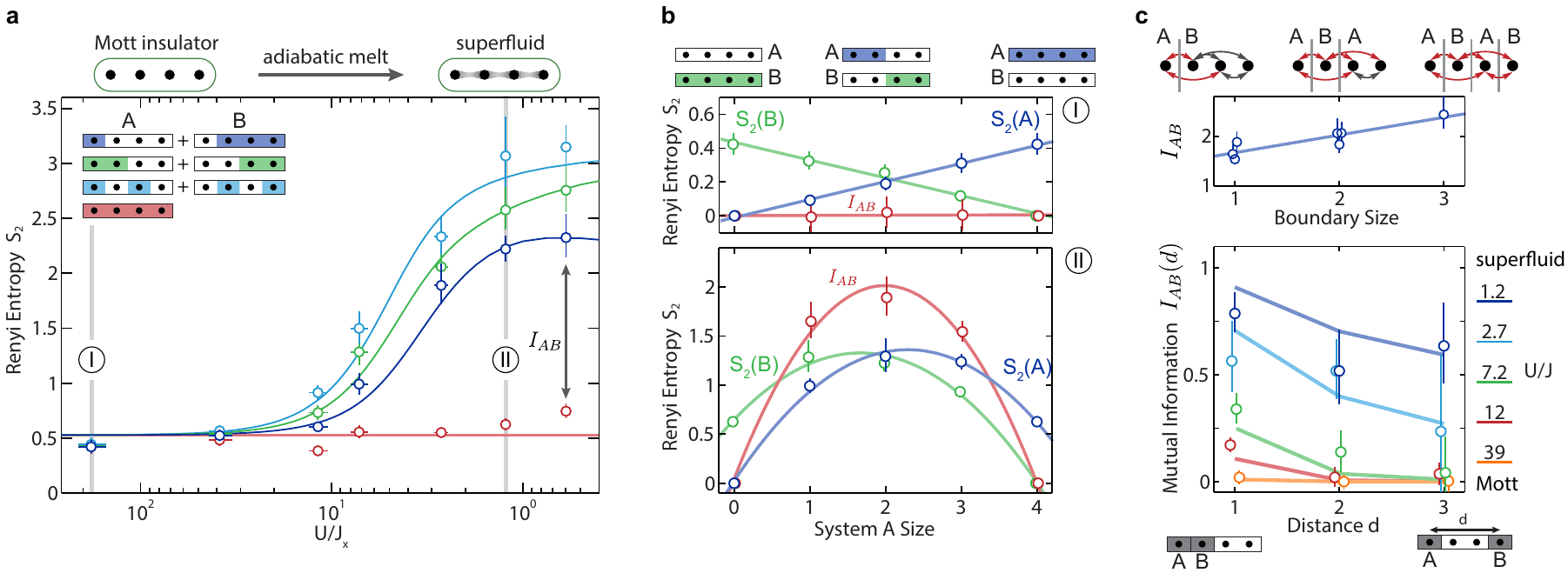}
\end{flushleft}

\caption{\label{fig:IAB}
{{\bf \Renyi Mutual information in the ground state.} Contribution from the extensive classical entropy in our measured \Renyi entropy can be factored out by constructing the mutual information $I_{AB}=S_{2}(A)+S_{2}(B)-S_{2}(AB)$. Mutual information takes into account all correlations \cite{Wolf2008} between the subsystems $A$ and $B$. {\bf a.} We plot the summed entropy $S_{2}(A)+S_{2}(B)$ (in blue, green and light blue corresponding to the partitions shown) and the entropy of the full system $S_{2}(AB)$ (in red) separately. Mutual information is the difference between the two, as shown by the arrow for a partitioning scheme. In the Mott insulator phase ($U/J_{x}\gg 1$) the sites are not correlated, and $I_{AB}\approx 0$. Correlations start to build up for smaller $U/J_{x}$, resulting in a non-zero mutual information. The theory curves are from exact diagonalization, with added offsets consistent with the extensive entropy in the Mott insulator phase. {\bf b.} Classical and entanglement entropies follow qualitatively different scaling laws in a many-body system. Top - In the Mott insulator phase classical entropy dominates and $S_{2}(A)$ and $S_{2}(B)$ follow a volume law-- entropy increases with the size of the subsystem. The mutual information $I_{AB}\approx 0$. Bottom - $S_{A}$, $S_{B}$ show non-monotonic behavior, due to the dominance of entanglement over classical entropy, which makes the curves asymmetric. $I_{AB}$ restores the symmetry by removing the classical uncorrelated noise. {\bf c.} Top - More correlations are affected (red arrow) with increasing boundary area, leading to a growth of mutual information between subsystems. The data points are for various partitioning schemes shown in Fig. \ref{fig:S2}b. Bottom- $I_{AB}$ as a function of the distance $d$ between the subsystems shows the onset and spread of correlations in space, as the Mott insulator adiabatically melts into a superfluid.}}
\end{figure*}

Experimentally, we find exactly this behavior for our two 4-site Bose-Hubbard systems (Fig. 4). We observe the emergence of spatial entanglement as the initial atomic Mott insulator melts into a superfluid. The measured quantum purity of the full system is about 0.6 across the Mott to superfluid crossover, corresponding to a \Renyi entropy, $S_{2}(AB)\approx 0.5$. The measured purity deep in the superfluid phase is slightly reduced, likely due to the reduced beam splitter fidelity in presence of increased single sites occupation number, and any residual heating. The nearly constant global purity indicates high level of quantum coherence during the crossover. For lower interaction strength $U/J_{x}$ (superfluid regime) we observe that the subsystem \Renyi entropy  is higher than the full system, $S_{2}(A)>S_{2}(AB)$. This demonstrates the presence of spatial entanglement in the superfluid state. In the Mott insulator regime ($U/J_{x}\gg 1$), $S_{2}(A)$ is lower than $S_{2}(AB)$ and proportional to the subsystem size, consistent with a product state. 

In these experiments, we post-select outcomes of the experiment for which the total number of atoms detected in both copies is even. This constitutes about 60\% of all the data, and excludes realizations with preparation errors, atom loss during the sequence, or detection errors (Supplementary material). The measured purity is consistent with an imperfect beam splitter operation alone, suggesting significantly higher purity for the many-body state. The measured entropy is thus a  sum of an extensive classical entropy due to the imperfections of the beam splitter and any entanglement entropy.

Our site resolved measurement simultaneously provides information about all possible spatial partitionings of the system. Comparing the purity of all subsystems with that of the full system enables us to determine whether a quantum state has genuine spatial multipartite entanglement where every site is entangled with each other. Experimentally we find that this is indeed the case for small $U/J_{x}$ (Fig. \ref{fig:S2}b). In the superfluid phase, all possible subsystems have more entropy than the full system, demonstrating full spatial multipartite entanglement between all four sites \cite{Alves2004, Palmer2005}. In the Mott phase ($U/J_{x}\gg 1$), the measured entropy is dominated by extensive classical entropy, showing a lack of entanglement.

By measuring the second-order \Renyi entropy we can calculate other useful quantities, such as the associated mutual information $I_{AB}=S_2(A)+S_{2}(B)-S_{2}(AB)$. Mutual information exhibits interesting scaling properties with respect to the subsystem size, which can be key to studying area laws in interacting quantum systems \cite{Wolf2008}. In some cases, such as in the `data hiding states' \cite{Divincenzo2002}, mutual information is more appropriate than the more conventional two point correlators which might take arbitrarily small values in presence of strong correlations.  Mutual information is also immune to present extensive classical entropy in the experiments, and hence is practically useful to experimentally study larger systems. In our experiments (Fig. \ref{fig:IAB}a), we find that for the Mott insulator state ($U/J_{x}\gg 1$), the entropy of the full system is the sum of the entropies for the subsystems. The mutual information $I_{AB}\approx 0$ for this state, consistent with a product state in the presence of extensive classical entropy. At $U/J_{x}\approx 10$, correlations between the subsystems begin to grow as the system adiabatically melts into a superfluid, resulting in non-zero mutual information, $I_{AB}>0$. 

It is instructive to investigate the scaling of \Renyi entropy and mutual information with subsystem size \cite{Wolf2008, Eisert2010} since in larger systems they can characterize quantum phases, for example by measuring the central charge of the underlying quantum field theory \cite{Calabrese2009}. Figure \ref{fig:IAB}b shows these quantities versus the subsystem size for various partitioning schemes with a single boundary. For the atomic Mott insulator the \Renyi entropy increases linearly with the subsystem size and the mutual information is zero, consistent with both a product state and classical entropy being uncorrelated between various sites. In the superfluid state the measured \Renyi entropy curves are asymmetric and first increase with the system size, then fall again as the subsystem size approaches that of the full system. This represents the combination of entanglement entropy and the linear classical entropy. This non-monotonicity is a signature of the entanglement entropy, as the entropy for a pure state must vanish when the subsystem size is zero or the full system. The asymmetry due to classical entropy is absent in the mutual information. 

The mutual information between two subsystems comes from the correlations across their separating boundary. For a four site system, the boundary area ranges from one to three for various partitioning schemes. Among those schemes with a single boundary maximum mutual information in the superfluid is obtained when the boundary divides the system symmetrically (Fig. \ref{fig:IAB}a). Increasing the boundary size increases the mutual information, as more correlations are interrupted by the partitioning (Fig.~\ref{fig:IAB}c).

Mutual information also elucidates the onset of correlations between various sites as the few-body system crosses over from a Mott insulator to a superfluid phase. In the Mott insulator phase ($U/J_{x}\gg 1$) the mutual information between all sites vanish (Fig. \ref{fig:IAB}c, bottom). As the particles start to tunnel only the nearest neighbor correlations start to build up ($U/J_{x}\approx 12$) and the long range correlations remain negligible. Further into the superfluid phase, the correlations extend beyond the nearest neighbor and become long range for smaller $U/J_{x}$. These results suggest disparate spatial behavior of the mutual information in the ground state of an uncorrelated (Mott insulator) and a strongly correlated phase (superfluid). For larger systems this can be exploited to identify quantum phases and the onset of quantum phase transitions.

\begin{figure}
\begin{center}
\includegraphics[width=89mm]{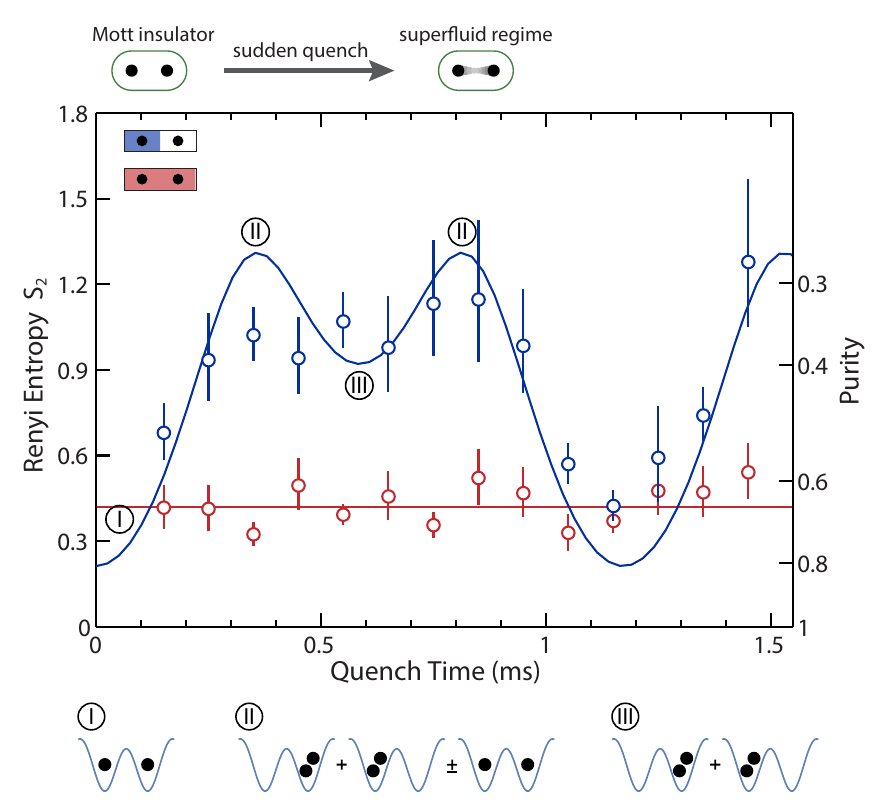}
\end{center}
\caption{\label{fig:quench}
{{\bf Entanglement dynamics in quench} Entanglement dynamics of two atoms in two sites after a sudden quench of the Hamiltonian from a large value of $U/J_{x}$ to $ U/J_{x}\approx 0.3$, with $J_{x} \approx 210$ Hz. Here, `quench time' refers to the duration that the atoms spend in the shallow double well, after the initial sudden quench. The system oscillates between Mott insulator like state (I) and quenched superfluid states (II, III). The growth of bipartite entanglement in the superfluid state is seen by comparing the measured \Renyi entropy of the single site subsystem (blue circles) to that of the two site full system (red circles). The solid lines are the theory curves with vertical offsets to include the classical entropy introduced by experimental imperfections.}}
\end{figure}

\section*{Non-equilibrium entanglement dynamics}
Away from the ground state, the non-equilibrium dynamics of a quantum many-body system is often theoretically intractable. This is due to the growth of entanglement beyond the access of numerical techniques such as the time dependent Density Matrix Renormalization Group (DMRG) theory \cite{Vidal2004, Trotzky2012}. 
Experimental investigation of entanglement may shed valuable light onto non-equilibrium quantum dynamics. Towards this goal, we study a simple system: two particles oscillating in a double well \cite{Kaufman2014, Trotzky2010}. This non-equilibrium dynamics are described by be Bose-Hubbard model. The quantum state of the system oscillates between unentangled (particles localized in separate wells) and entangled states in the Hilbert space spanned by $|1,1\rangle$, $|2,0\rangle$ and $|0,2\rangle$.  Here, $|m,n\rangle$ denotes a state with $m$ and $n$ atoms in the two subsystems (wells), respectively. Starting from the product state $|1,1\rangle$ the system evolves through the maximally entangled states $|2,0\rangle+|0,2\rangle\pm |1,1\rangle$ and the symmetric HOM-like state $|2,0\rangle+|0,2\rangle$. In the maximally entangled states the subsystems are completely mixed, with a probability of $1/3$ to have zero, one, or two particles. The system then returns to the initial product state $|1,1\rangle$ before re-entangling. In our experiment, we start with a Mott insulating state ($U/J_{x}\gg 1$), and suddenly quench the interaction parameter to a low value, $U/J_{x}\approx 0.3$. The non-equilibrium dynamics is demonstrated (Fig. \ref{fig:quench}) by the oscillation in the second-order \Renyi entropy of the subsystem, while the full system assumes a constant value originating from classical entropy. This experiment also demonstrates entanglement in HOM-like interference of two massive particles.

\section*{Summary and outlook}

In this work, we perform a direct measurement of quantum purity, the second-order \Renyi entanglement entropy, and mutual information in a Bose-Hubbard system. Our measurement scheme does not rely on full density matrix reconstruction or the use of specialized witness operators to detect entanglement. Instead, by preparing and interfering two identical copies of a many-body quantum state, we probe entanglement with the measurement of only a single operator. Our experiments represent an important demonstration of the usefulness of the many-body interference for the measurement of entanglement. It is straight forward to extend the scheme to fermionic systems \cite{Pichler2013} and systems with internal degrees of freedom \cite{Alves2004}. By generalizing the interference to $n$ copies of the quantum state \cite{Brun2004}, arbitrary observables written as as $n$-th order polynomial function of the density matrix, e.g. $n>2$ order \Renyi entropies, can be measured.

With modest technical upgrades to suppress classical fluctuations and residual interactions, it should be possible to further improve the beam splitter fidelity enabling us to work with significantly larger systems. Mutual information is an ideal tool for exploring these larger systems as it is insensitive to any residual extensive classical entropy. For high entropy of a sub-system, corresponding to low state purity, the number of measurements required to reach a desired precision is high.  However, in contrast to tomographic methods, this scheme would not require additional operations for larger systems. Moreover, the single site resolution of the microscope allows us to simultaneously obtain information about all possible subsystems, to probe multipartite entanglement.

For non-equilibrium systems, entanglement entropy can grow  in time (indefinitely in infinite systems). This leads to interesting many-body physics, such as thermalization in closed quantum systems \cite{Rigol2008}. The long time growth of entanglement entropy is considered to be a key signature of many-body localized states \cite{Bardarson2012} arising in presence of disorder. The ability to measure the quantum purity for these systems would allow experimental distinction of quantum fluctuations and classical statistical fluctuations. 

More generally, by starting with two different quantum states in the two copies this scheme can be applied to measure the quantum state overlap between them. This would provide valuable information about the underlying quantum state. For example, the many-body ground state is very sensitive to perturbations near a quantum critical point. Hence, the overlap between two ground states with slightly different parameters (such as $U/J$ in the Bose-Hubbard hamiltonian) could be used as a sensitive probe of quantum criticality \cite{zanardi2006}. Similarly the overlap of two copies undergoing non-equilibrium evolution under different perturbations can be used to probe temporal correlation functions in non-equilibrium quantum dynamics.

We thank J. I. Cirac, M. Cramer, A. Daley, A. DelMaestro, M. Endres, S. Gopalakrishnan, A. Kaufman, M. Knap, A. Pal, H. Pichler, B. Swingle, and P. Zoller for useful discussions. Supported by grants from the Gordon and Betty Moore FoundationÕs EPiQS Initiative (grant  GBMF3795), NSF through the Center for Ultracold Atoms, the Army Research Office with funding from the DARPA OLE program and a MURI program, an Air Force Office of Scientific Research MURI program, and an NSF Graduate Research Fellowship (M.R.).

\bibliography{HOM_arXiv}

\setcounter{equation}{0}

\renewcommand{\theequation}{S.\arabic{equation}}

\section*{Supplementary Material}

\section{Measuring entanglement entropy with quantum interference}

The quantification of entanglement requires the measurement of non-linear functionals of a quantum state $\rho$, such as the $n$-th order R\'{e}nyi entropy $S_{n}=-\ln \mathrm{Tr}(\rho^n)$ \cite{mintert2007s}. A general scheme to measure $\mathrm{Tr}(\rho^n)$ is to measure the shift operator $V_{n}$ acting on $n$-copies of the many-body system. The shift operator $V_n$ re-orders the quantum states when acting on a collection of $n$ states,
\begin{equation}
V_n |\psi_1\rangle |\psi_2\rangle \dots |\psi_n\rangle = |\psi_n\rangle |\psi_1\rangle \dots |\psi_{n-1}\rangle.
\end{equation}
It can be shown that $\mathrm{Tr}(\rho^n) = \mathrm{Tr}(V_n \rho^{\otimes n}) $ \cite{ekert2002s}.

We focus on the experimentally relevant case of $n=2$. The shift operator is then simply the SWAP operator which exchanges any two quantum states:
\begin{equation}
V_2 ( |\psi_1\rangle  \otimes |\psi_2\rangle )  =  |\psi_2\rangle  \otimes |\psi_1\rangle
\end{equation}
Two successive applications of the SWAP operator leave the system unchanged, $V_2^2 = \mathbb{1}$. Therefore $V_2$ has eigenvalues $\pm 1$, corresponding to subspaces of the $2$-copy system that are symmetric or antisymmetric with respect to the state exchange. The SWAP operator may act on individual modes (e.g. lattice sites) or the entire quantum system, and operations on different modes commute. The following short proof\cite{daley2012s}
 \begin{align}
 \rm{Tr}\left(V_2 \rho_1 \otimes \rho_2 \right)  & = \rm{Tr}\left(  V_2 \sum_{ijkl} \rho_{ij}^{(1)} \rho_{kl}^{(2)} \left| i\right\rangle \left\langle j\right| \otimes \left| k\right\rangle \left\langle l\right| \right)\nonumber  \\
 & = \rm{Tr}\left( \sum_{ijkl} \rho_{ij}^{(1)} \rho_{kl}^{(2)} \left| k\right\rangle \left\langle j\right| \otimes \left| i\right\rangle \left\langle l\right| \right) \nonumber  \\
  & = \sum_{ijkl} \rho_{ij}^{(1)} \rho_{kl}^{(2)} \delta_{kj} \delta_{il}  = \sum_{ik} \rho_{ik}^{(1)} \rho_{ki}^{(2)} = \rm{Tr}\left( \rho_1 \rho_2 \right)
 \end{align}
shows that the overlap of two quantum states $\rho_1$ and $\rho_2$ is given by the expectation value of the SWAP operator on the product space of the two states. Consider from now on the case where the two state are identical ($\rho_1=\rho_2=\rho$), then the expectation value of $V_2$ gives the purity $\tr{(\rho^{2})}$. Further if we have two copies of a pure state then $\tr{(\rho^{2})} = 1 = \tr{(V_2\rho\otimes\rho)}$, hence the combined $2$-copy state is symmetric and can be expressed in the symmetric basis comprised of states
\begin{equation}
\left\lbrace \prod_{i,j} (a_{1,i}^{\dagger} - a_{2,i}^{\dagger})^{2p_i} (a_{1,j}^\dagger + a_{2,j}^\dagger)^{q_j} \left| \textrm{vac}\right\rangle 	\right\rbrace \hspace{0.1in}\textrm{with}\hspace{0.1in} p,q=0,1,2,\dots
\label{eq:symmetricbasis}
\end{equation}
where $a_{1(2),i}^{\dagger}$ is the creation operation of mode $i$ in copy $1$($2$). If the two copies undergo a discrete Fourier transformation of the form (for simplicity dropping the mode indices)
 \begin{align}
 (a^{\dag}_1 +  a^\dag_{2})/\sqrt{2}  & \rightarrow a^\dag_1 \nonumber\\
 (a^\dag_{1} - a^\dag_{2})/\sqrt{2} & \rightarrow  a^\dag_{2} 
 \label{fourier}
 \end{align}
then the basis states in Eq.\ref{eq:symmetricbasis} will end up having $2\times p_i$ particles in mode $i$ of copy $2$. In other words a symmetric state, as is the case for 2 pure identical copies, will always have \textit{even} number of particles in copy $2$ after the transformation. The symmetric and anti-symmetric subspaces of the SWAP operator are identified by the parity of atom number in copy $2$ after a discrete Fourier transform, and the average parity directly measures the state purity, $\left\langle P_2\right\rangle = \tr{\rho}^{2}$. 

Our microscope experiments then allows us to probe entanglement in an optical lattice by comparing the local purity $\tr{\rho_{A(B)}}^{2}$ to the global purity $\tr{\rho}^{2}$ for a system partitioned into subsystems $A$ and $B$. The entanglement is quantified by the entropy of the reduced density matrix $\rho_A = \mathrm{Tr}_B(\rho)$, and the measured purity $\tr{\rho^{2}}$ gives directly the 2nd order R\'{e}nyi entropy $S_{2}=-\ln{\tr{\rho}^{2}}$. This scheme is proposed in \cite{ekert2002s} and made explicit for measurements with beamsplitter operations in optical lattices in \cite{Alves2004s} and \cite{daley2012s}, giving R\'{e}nyi entropy of arbitrary order $n$.\\

\begin{figure}
	\begin{center}
		\includegraphics[width=1 \linewidth]{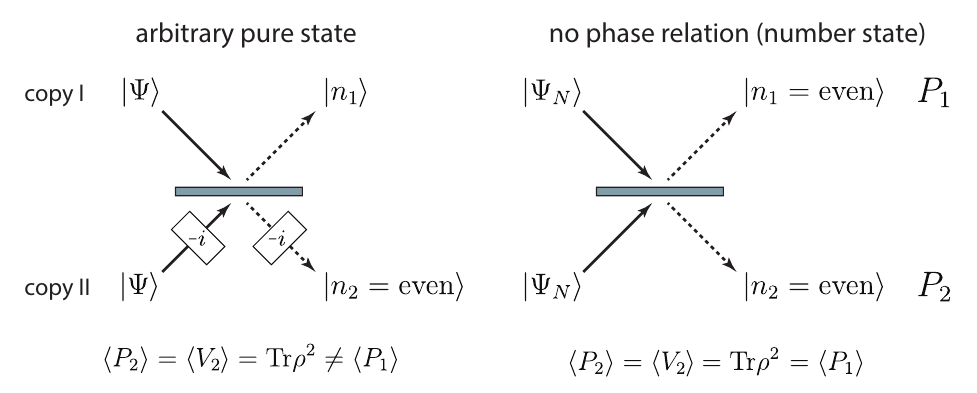}
		\caption{\textbf{Beamsplitter for many-body interference.} \textit{Left:} With the
		beamsplitter operation and proper phase shift operations, one can directly measure
		quantum purity by measuring the average parity in output port~2 of the beamsplitter.
		For pure identical incident states, the atom number is always even in output~2.
		\textit{Right:} In the experiment, we interfere states with well-defined particle
		number $N$ or subsystems of such states. No macroscopic phase relationship exists
		between the input states, and the phase shifts in the input/output ports have no
		physical significance. Both outputs are equivalent and may be used to measure the
		expectation value $\langle V_2 \rangle$ of the SWAP operator. }
		\label{fig:phasefactor}
	\end{center}
\end{figure}

Using controlled tunneling in a double-well potential, we can implement the beamsplitter transformation for bosonic atoms (see next section):
 \begin{align}
 a^\dag_1  & \rightarrow ( a^\dag_1 + i a^\dag_{2})/\sqrt{2} \nonumber\\
 a^\dag_{2} & \rightarrow ( i a^\dag_{1} + a^\dag_{2})/\sqrt{2}
 \label{beamsplitter}
 \end{align}
 where a $\pi/2$-phase ($i$) is associated with each tunneling event across the double-well. Note that this transformation is \textit{not} equivalent to the Fourier transform in Eq. (\ref{fourier}). It's easy to verify that the Fourier transform is realized with the following protocol sequence of the beamsplitter operation and relative phase shift operations:
 
 \begin{enumerate}
 	\item A $-\pi/2$ phase shift ($-i$) on copy 2
 	\item The beamsplitter operation in Eq. (\ref{beamsplitter})
 	\item Another $-\pi/2$ phase shift on copy 2
 \end{enumerate}

The inclusion of the additional phase shifts are important to correctly map the
symmetric (antisymmetric) eigenstates of the SWAP operator onto states with even
(odd) atom number parity in output port 2 of the beamsplitter. The resulting
protocol is valid for measuring purity of any general many-body state. In the
classical limit where the incident states are two identical coherent states with
well-defined identical phases, the inclusion of the proper phase factors in input 2
ensures that the states interfere destructively in output 2. In this port, the total
number of bosons is always zero and therefore even, so the measured parity $\langle
V \rangle = 1$ correctly gives $\rho_1 = \rho_{2}$ and $\tr{\rho}^{2}=1$. This
situation is analogous to the interference of two phase-stabilized laser beams on a
50/50 beamsplitter, which may result in zero intensity in one output for the correct
choice of incident phases. Away from the classical limit, for example as the input
states become number squeezed states with decreasing uncertainty in atom number but
increasing fluctuation in phase, atoms start to appear in output port 2 after the
protocol but only in pairs (even parity) as long as the input states remain pure and
identical.

The protocol also works when there is no global phase relationship between the
interfering many-body states. Such as in our current experiments when the two copies
are prepared each as an independent quantum state with a fixed number of
atoms, so there is no well-defined phase. There are also no defined phases when the
incident states to the beamsplitter are subsystems partitioned out of bigger
systems. In either case, step $1$ of the above protocol has no physical meaning in
the absence of a defined phase and might be omitted from the experiment without
changing the resulting state after the transformation. The \textit{in-situ} fluorescence
imaging of our microscope detects parity of the atom number on each lattice site
which is phase-insensitive, so step $3$ is also redundant. The beamsplitter
operation in the double-well alone is thus sufficient to implement the mapping of
SWAP operator eigenstates onto states with even or odd atom number parity. The two
output ports are then equivalent and the purity measurements may be obtained from
the atom numbers on either side of the double-well after the many-body interference
sequence.\\

\section{Implementation of the beamsplitter operation}

\subsection{Projected double-well potentials} 
In addition to a square lattice, optical potentials are generated by projecting light onto the atoms using a digital micro-mirror device (DMD). The DMD is used as an amplitude hologram in a Fourier plane of our high resolution imaging system so that wavefronts with arbitrary phase and amplitude may be created with single-site resolution \cite{zupancic2013s}. We use blue-detuned coherent light at $\lambda = 762$ nm to generate a potential with a double-well profile along $x$ and a smoothed flat top profile along $y$:

\begin{eqnarray}
V(x,y)&=& V_{dw} \left (e^{-\frac{(x-1.5)^2}{0.95^2}}-0.52 e^{-\frac{x^2}{0.9^2}}+e^{-\frac{(x+1.5)^2}{0.95^2}} \right )^2 \nonumber \\
& &\times \left( \arctan \left( \frac{y+18}{5.5} \right) - \arctan \left( \frac{y-18}{5.5} \right) \right)^2\nonumber\\
\label{eqn:Vxy}
\end{eqnarray}

where $x$ and $y$ are in units of the lattice spacing and $V_{dw}$ is the potential depth of the projected double-well. 

\begin{figure}
\begin{center}
\includegraphics[width=0.9 \linewidth]{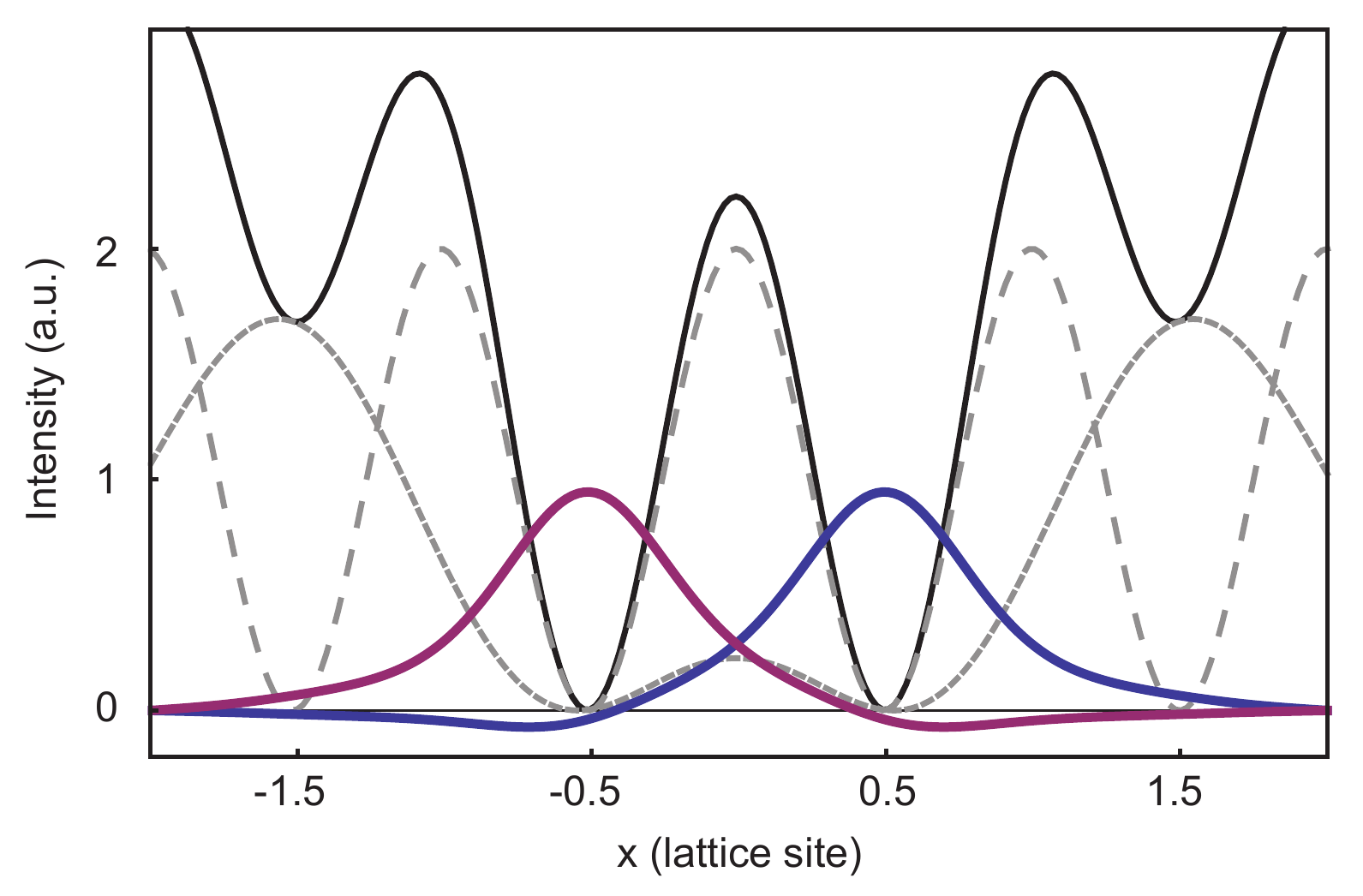}
\caption{\textbf{Double-well potential for the beamsplitter.} The intensity profile of the projected potential (Eq.\ref{eqn:Vxy}, gray short-dashed), the lattice (gray long-dashed), and the combined potential for the beam splitter operation (black solid). Also shown are sketches of the amplitude of the ground band Wannier wavefunctions (blue, purple) in each well at the beamsplitter depth.}
\label{fig:doublewell}
\end{center}
\end{figure}

\begin{figure}
\begin{center}
\includegraphics[width=0.9\linewidth]{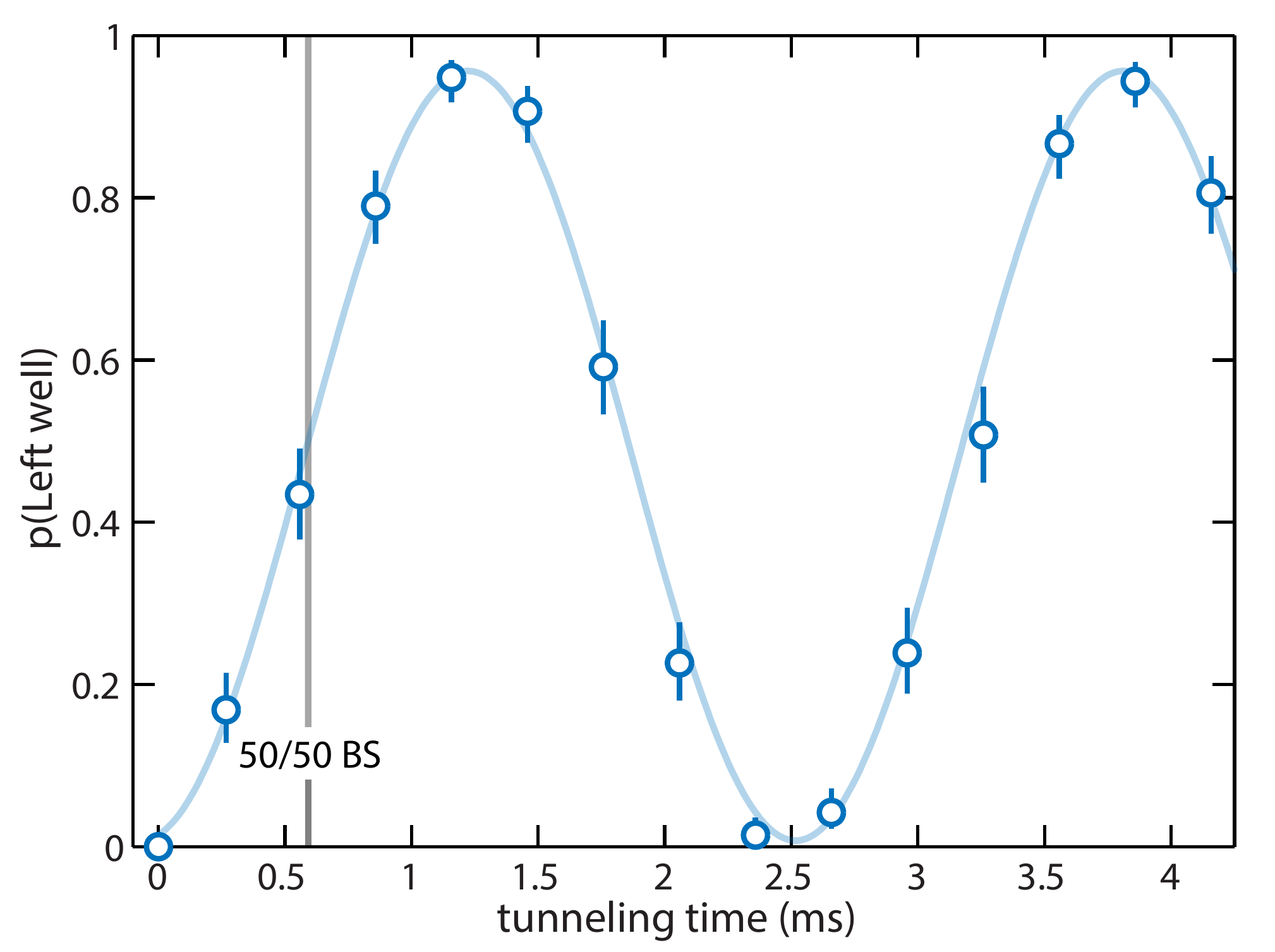}
\caption{\textbf{Rabi oscillations in the double-well.} A single particle is initialized on the right side of the doublewell and oscillates coherently between the two wells with fitted tunneling rate $J=193(4)$\,Hz and contrast of $95(1)\%$.}
\label{fig:rabi}
\end{center}
\end{figure}

The beamsplitter operation is realized by controlled tunneling in the combined potential of the above projected potential and a shallow $x$-lattice, as depicted in Fig.\,\ref{fig:doublewell}. We choose depths $V_{dw} = 1.7\,E_r$ and $V_{latt} = 2\,E_r$, for which we observe tunneling rate $J=193(4)$\,Hz during the beam-splitter operation (Fig.\,\ref{fig:rabi}), in reasonable agreement with a band structure calculation predicting $J=170$\,Hz. The discrepancy is likely due to uncertainly in the lattice depth, which is calibrated using amplitude modulation spectroscopy at $V_{latt} \approx 40\,E_r$. Here $E_{R}=1240$ Hz is the recoil energy of the optical lattice. In the beamsplitter potential, the energy gap to the first excited band is $\approx$ 1.3\,kHz, and states outside the ground band do not contribute significantly to the dynamics. 

\subsection{Sources of error for the beamsplitter} 

\begin{figure}
\begin{center}
\includegraphics[width=0.9\linewidth]{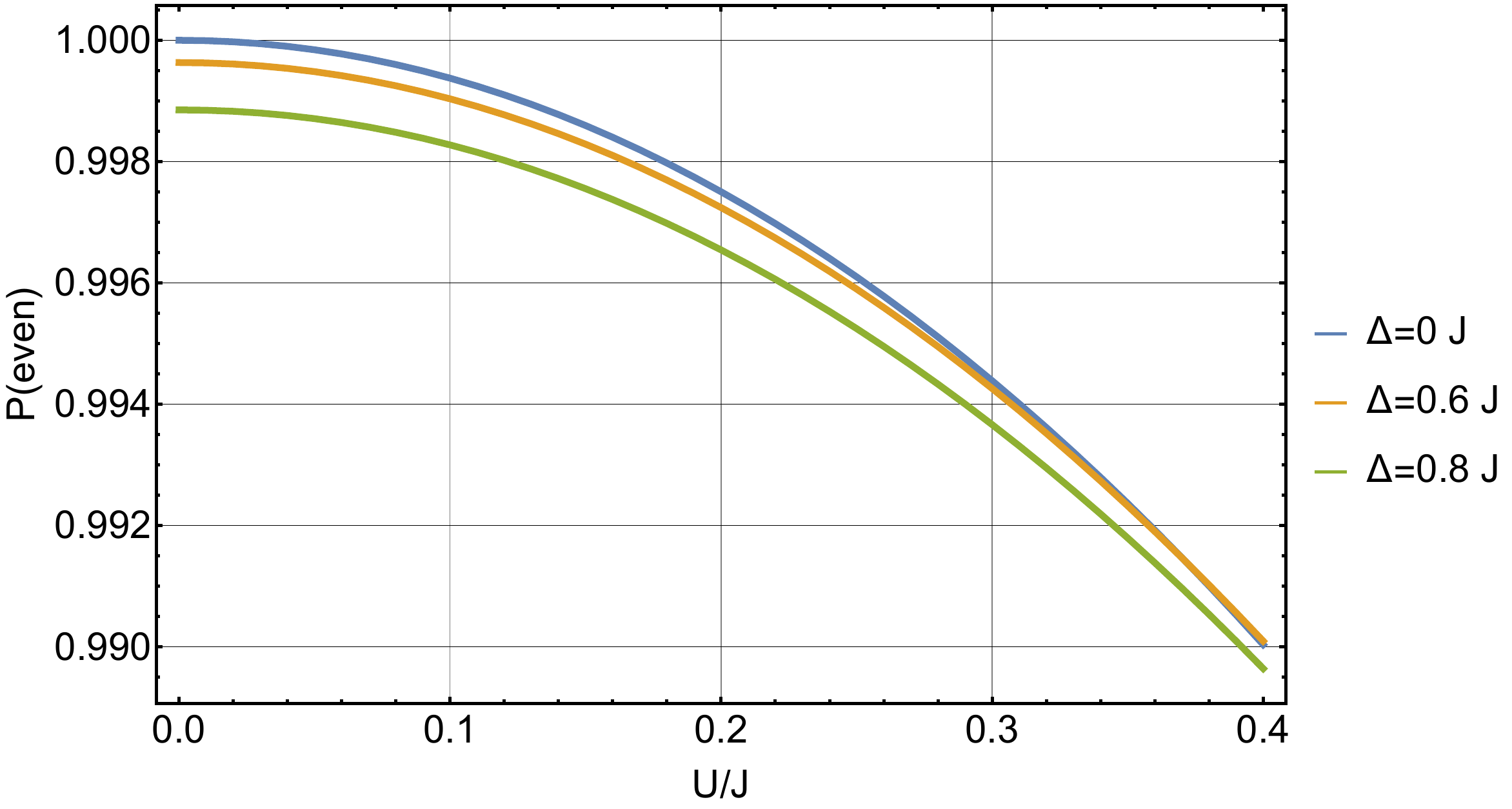}
\caption{\textbf{Fidelities of the beamsplitter operation.} Finite interactions and energy offsets due to imperfections in the double-well potential reduce the Hong-Ou-Mandel interference contrast, as measured by the probability to detect even atom numbers at the beamsplitter time $tJ=\frac{\pi}{4}$. For a beamsplitter operation starting with one atom on each side of the double-well and typical experimental parameters $J=240$\,Hz, $U=70$\,Hz and offset $\Delta=50$\,Hz (corresponding to $U/J=0.3$ and $\Delta/J=0.2$ ), the interference contrast is expected to be reduced by $\sim 0.6\%$. This calculation does not take the effects of higher bands into account.}
\label{fig:HOM_contrast}
\end{center}
\end{figure}

\paragraph{Potential imperfections} The leading order imperfection of the projected double-well potential are imperfect zero-crossings in the electric field, resulting in energy offsets between the two sides of the double-well. At the double-well depth for our beamsplitter operation, we observe offsets of 50\,Hz or less, which do not significantly affect the Hong-Ou-Mandel (HOM) interference contrast (see Fig.~\ref{fig:HOM_contrast}).

\paragraph{Alignment stability} The successful loading of atoms from the lattice into the double-well potential is sensitive to long-term and shot-to-shot position drifts between the lattice and the double-well. We minimize such drifts by imaging the lattice and double-well potential at the end of each experimental run and feeding back on the position of the double-well with a piezo-actuated mirror. We achieve a relative position stability of 0.04 sites RMS or less. To lowest order the position drift creates an energy offset between the two sides of the combined double-well potential. At the chosen depths for the beamsplitter operation, a relative shift of 0.1 sites leads to an offset of $\approx$ 20\,Hz.

\paragraph{Interaction effects}
Interactions during the beamsplitter operation potentially reduce the HOM interference contrast. We minimize interactions by performing all experiments in a weak out-of-plane confinement of $\omega_z=2\pi \times 800$\,Hz. During the beamsplitter operation we achieve an interaction of $U=2\pi \times 70$\,Hz (measured with photon-assisted tunneling in a deep double-well and extrapolated to lower depths), corresponding to $u=U/J=0.3$. This residual interaction reduces the HOM interference contrast by $\sim 0.6\%$ (see Fig.~\ref{fig:HOM_contrast}).

\paragraph{Coherent admixture of higher bands}Interactions of two particles on the same site distort the particles' wavefunctions and coherently admix higher bands. This wavefunction is thus different from that of a single particle, restoring some distinguishability to the bosonic atoms. The dominant contribution of higher bands occurs in the $z$-direction, along which the confinement is weakest, and the second excited band is admixed to the wavefunction. The admixture is $\epsilon \approx (\frac{U}{2 \omega_z})^2=(\frac{2 \pi \times 70\mathrm{Hz}}{2\times 2 \pi \times 800\mathrm{Hz}})^2=0.2$\%. HOM interference contrast is thus reduced by less than $1\%$.


\section{Experimental Sequence}

\begin{figure}
\begin{center}
\includegraphics[width=0.95\linewidth]{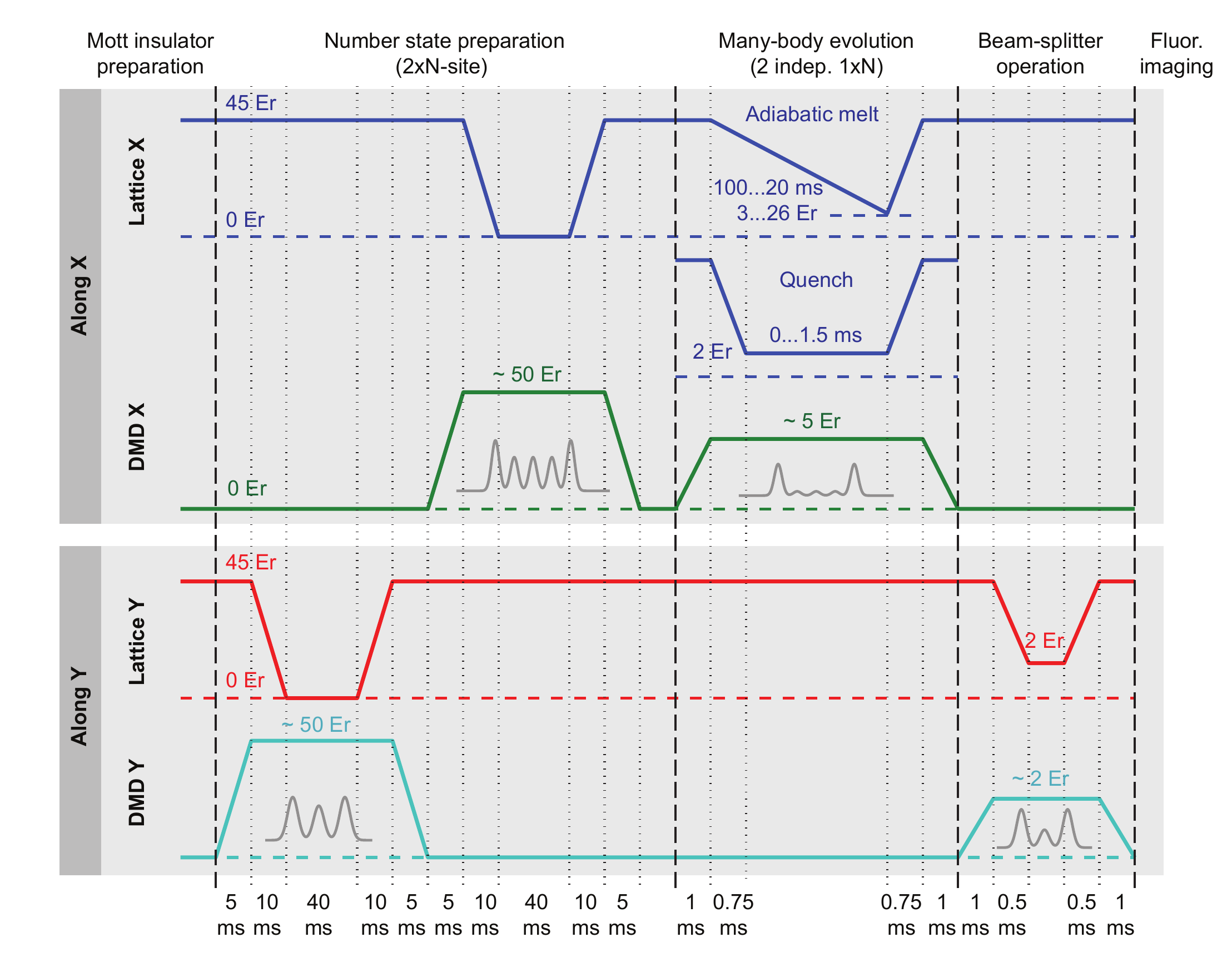}
\caption{\textbf{Experimental sequence.} Schematic showing the ramps of the $x$- and $y$- lattices and the projected potential from the DMD. The profiles of the DMD potentials are sketched in the direction of interest, while the other direction always has a smoothed flattop profile aross the region of interest. Ramps are exponential in depth as a function of time. See text for details.}
\label{fig:sequence}
\end{center}
\end{figure}

Our experiments start with a single-layer two-dimensional Mott insulator of $^{87}$Rb atoms in a deep lattice ($V_x=V_y=45 E_r$) with $680$\,nm spacing as described in previous work. The following sequence is illustrated in Fig.\ref{fig:sequence}.

\paragraph*{State preparation} 
We deterministically prepare a plaquette of $2 \times 2$ or $2 \times 4$ atoms from a Mott insulator with unity occupancy. We first superimpose onto the deep lattice an optical potential with a double-well profile of depth $\approx 50 E_r$ along $y$ and a smooth flattop profile along $x$, and subsequently turn off the lattice in the $y$-direction. The two troughs of the double-well are aligned with two neighboring lattice sites so only two rows of atoms are trapped, while all other atoms leave the system along tubes in the $y$-direction. A blue-detuned Gaussian beam with waist $w \approx 50\,\mu$m and $\lambda = 755$\,nm provides the anti-confinement to remove atoms outside the double-well efficiently in 40 ms. The y-lattice is then ramped back to its initial depth and the double-well removed, leaving a block of width 2 sites and length $\sim$ 10 sites populated with one atom each. The above procedure is then repeated with a double- or quadruple-well potential along the $x$ direction, leaving a deterministically loaded block of $2 \times 2$ or $2 \times 4$ atoms in the lattice. The lattices and double-well potentials are ramped smoothly to avoid heating the atoms to excited bands of the lattice.

At the end of the state preparation sequence, the fidelity of unit occupancy is $94.6(2)\%$ per site, limited primarily by the fidelity of the initial Mott insulator and losses during the state preparation. We verify independently that defects are predominantly empty, not doubly occupied sites. 

\paragraph*{Evolution in independent copies} 

For studying the ground state entanglement using the $2 \times 4$ block (Figure. 4 \& 5 in the main text), we turn on an optical potential with two narrow Gaussian peaks separated by four lattice sites along the $x$ direction and flat-top along $y$. This confines the atoms inside the 4-site ''box potential''. The $x$ lattice is then ramped down adiabatically to various final depths from $26$ to $3 E_r$. The ramp in depth is exponential in time with a duration of 200\,ms from $45$ to $3 E_r$. The $y$-lattice is kept at $45 E_r$ so that tunneling along $y$ is negligible and the two copies evolve independently. 

For quench dynamics using the $2 \times 2$ block, we use a double-well potential along $x$ with $V_{dw}=2 E_r$ to prevent atoms from leaving the 2-site system. The $x$ lattice is ramped from $45 E_r$ to $2 E_r$ in $0.75$\,ms and held for a variable time. The presence of the double-well slightly modifies the values of $U$ and $J$ from values in a lattice only, yielding $U/J = 0.3$ during the hold time.

\paragraph*{Beamsplitter operation and imaging} 
Right before the beamsplitter operation, the $x$-lattice is ramped back to $45 E_r$ in 0.75\,ms to suppress tunneling within each copy. A double-well potential along $y$ is superimposed onto the lattice. The $y$-lattice is then ramped down to $2 E_r$ in 0.5\,ms and atoms are allowed to tunnel in independent double-wells between the two copies for 0.34\,ms, implementing the beam splitter transformation before the $y$-lattice is returned to its initial depth of $45 E_r$ in 0.5\,ms.

Subsequently, we pin the atoms in a deep lattice and obtain fluorescence images with single-lattice-site resolution. Our detection scheme is sensitive only to the parity of the site occupation number.

\begin{figure*}

\begin{center}
\includegraphics[width=0.85 \linewidth]{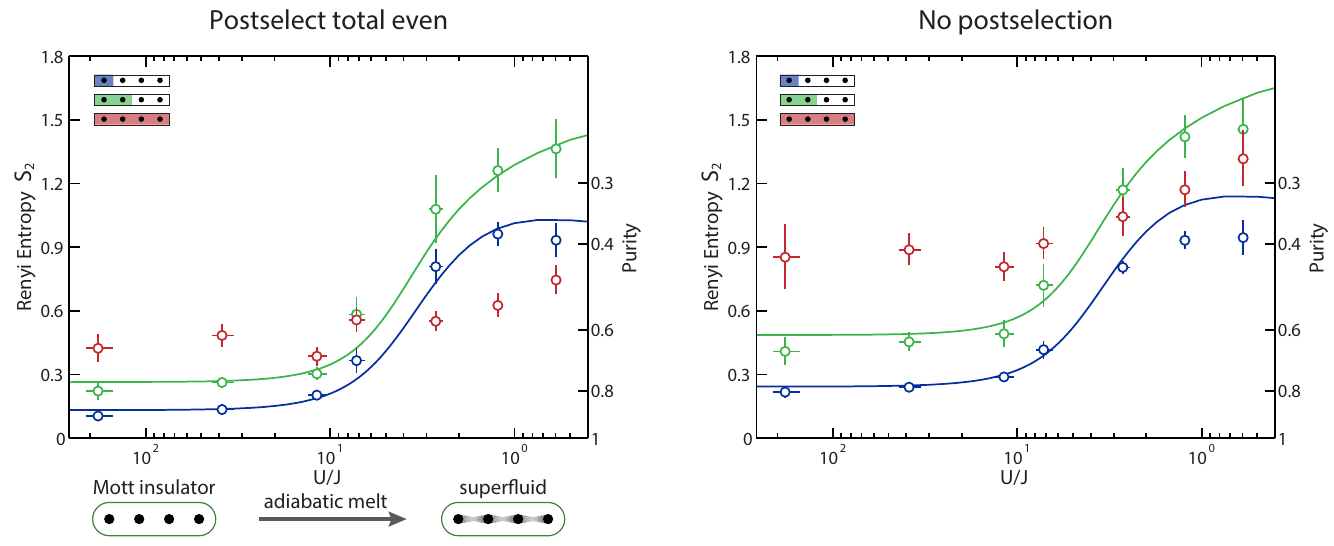}
\end{center}
\caption{\textbf{R\'{e}nyi entropy of the 4-site system and its subsystems with and without postselection.} The postselection process removes classical entropy and reduces the entropy of the full system (red) from $\sim 0.9$ to $\sim 0.5$. Note that even without postselection the entropy of the half-chain (green) rises above the full system entropy, indicating bipartite entanglement. Theory curves are exact diagonalizations shifted vertically by the mean classical entropy per site calculated from the full system entropy.}
\label{fig:postselection_fig}
\end{figure*}

\section{Data Analysis}
\paragraph*{Post-selection} Before data analysis we post-select outcomes of the experiment for which the total number of atoms detected within the plaquette ($2 \times 2$ or $2 \times 4$ sites) is even. Outcomes outside this subset of data indicate either state preparation errors, atom loss during the sequence, or detection errors. We furthermore reject all realizations for which we detect atoms in the $20 \times 20$ block surrounding the region of interest, most likely corresponding to atoms being lost from the plaquette during the sequence. Note that a combination of multiple errors (e.g. failure to load atoms on two sites) may lead to an unsuccessful run of the experiment being included in the post-selected data.\\

For experiments on the $2 \times 2$ plaquette we typically retain $80\%$ of the data, and $60\%$ for the  $2 \times 4$ plaquette.

\paragraph*{Calculating Purity and Entropy}

For the full many-body state and each subsystem of interest we calculate $p_\mathrm{odd}$, the probability of measuring a total odd number of atoms after the beamsplitter operation within the post-selected data. The quantum mechanical purity and second-order R\'{e}nyi entropy are then given by

\begin{align}
\mathrm{Tr}(\rho^2)&=1-2 p_\mathrm{odd}\\
S_2(\rho)& = - \log( \mathrm{Tr}(\rho^2))
\end{align}

We average the calculated purity $\mathrm{Tr}(\rho^2)$ over both copies and over equivalent partitions. For instance, the single-site entropy reported in Fig. 4a of the main text is the mean over the first and last site of each copy of the 4-site system. From the variance of the parity in each subsystem and the covariance between subsystems we obtain the statistical standard error of the mean parity, taking into account possible correlations between regions. The reported error bars are the quadrature sum of the statistical error and the standard deviation of mean parities over the averaged regions. This procedure accounts for residual inhomogeneities between the copies and along the chains.

Errorbars in $U/J$ correspond to a typical uncertainty in the optical lattice depth of $\pm 2 \%$.

\paragraph*{Full system purity}
For the $2 \times 4$ plaquette, the initial state purity is reduced from 1 due to the presence of thermal holes in the initial Mott insulating state. Assuming all even sites are holes, the loading statistics for the $2 \times 4$ plaquette are:\\

\begin{center}
\begin{tabular}{|c|c|}
\hline $N$ atoms & loading probability $p(N)$ \\ 
\hline 8 & 0.66(1) \\ 
\hline 7 & 0.27(1) \\ 
\hline 6 & 0.052(4)\\ 
\hline 
\end{tabular} 
\end{center}

i.e. the postselected subset of total even data contains $\frac{0.052}{0.052+0.66}=7\% $ of events with 6 atoms total. The inclusion of outcomes with 6 atoms reduces the purity of the initial state to 0.94, corresponding to a R\'{e}nyi entropy of 0.06.
The expected full system purity in the Mott insulator state is thus limited by the finite $95(3)\%$ fidelity of the beamsplitter operation on each site and approximately given by the product of individual purities, $P = \prod_k p^{(k)} = 0.90^4 \approx 0.66 $, in good agreement with the experimentally measured purity in Fig.~4a.

\paragraph*{Fitting procedure} To determine the contrast of single-particle Rabi oscillations (Fig.\ref{fig:rabi}) and HOM-interference (Fig.~3b in main text) we use a Bayesian inference for the fit to the measured parity, which is more robust than a least-squares fit in situations where error probabilities are small and the visibility close to 1. This approach prevents unphysical fits that extend past the possible bounds of the measurement and appropriately adjusts the error bars for points measured to lie near the physical bound. For each time point, we report the median and the $1 \sigma $(68\%) confidence interval of a Beta-distribution $\beta(p,m,N)$ for m successful outcomes in N experimental runs. The fitted sine curves in Fig. 1 are maximum-likelihood fits to the Beta-distributions at each time point, which are determined by maximizing the product of all the Beta-distributions where the fitted curve samples them. \cite{scheel09}


\end{document}